\documentclass[twocolumn,preprintnumbers,amsmath,amssymb,superscriptaddress]{revtex4}

\usepackage{graphicx}
\usepackage{dcolumn}
\usepackage{bm}
\usepackage{soul}
\usepackage{color}
\usepackage{epstopdf}
\usepackage{lipsum}
\usepackage[outercaption]{sidecap}
\usepackage{floatrow}
\begin{document}

\title{Theoretical model for the structural relaxation time in co-amorphous drugs}

\author{Anh D. Phan}
\affiliation{Faculty of Materials Science and Engineering, Phenikaa Institute for Advanced Study, Phenikaa University, Hanoi 100000, Vietnam}
\affiliation{Faculty of Information Technology, Artificial Intelligence Laboratory, Phenikaa University, Hanoi 100000, Vietnam}
\email{anh.phanduc@phenikaa-uni.edu.vn}
\author{Justyna Knapik-Kowalczuk}
\affiliation{Institute of Physics, University of Silesia, SMCEBI, 75 Pułku Piechoty 1a, 41-500 Chorzów, Poland}
\author{Marian Paluch}
\affiliation{Institute of Physics, University of Silesia, SMCEBI, 75 Pułku Piechoty 1a, 41-500 Chorzów, Poland}
\author{Trinh X. Hoang}
\affiliation{Institute of Physics, Vietnam Academy of Science and Technology, 10 Dao Tan, Ba Dinh, Hanoi, Vietnam}
\author{Katsunori Wakabayashi}
\affiliation{Department of Nanotechnology for Sustainable Energy, School of Science and Technology, Kwansei Gakuin University, Sanda, Hyogo 669-1337, Japan}
\date{\today}

\begin{abstract}
We propose a simple approach to investigate the structural relaxation time and glass transition of amorphous drugs. Amorphous materials are modeled as a set of equal sized hard spheres. The structural relaxation time over many decades in hard sphere fluids is theoretically calculated using the Elastically Collective Nonlinear Langevin Equation theory associated with Kramer's theory. Then, a new thermal mapping from a real material to an effective hard sphere fluid provides temperature-dependent relaxation time, which can compare to experiments. Numerical results quantitatively agree with previous experiments for pharmaceutical binary mixtures having different weight ratios. We carry out experiments to test our calculations for an ezetimibe-simvastatin-Kollidon VA64 mixture. Our approach would provide a simple but comprehensive description of glassy dynamics in amorphous composites. 
\end{abstract}

\maketitle
\section{Introduction}
Amorphous drugs have received considerable attention in recent years since their enhanced dissolution rate, water solubility, and bioavailability exceed that of their crystalline counterparts \cite{25,26,19,20,21}. The drugs lack long-range molecular ordering similar to liquids but have rheological properties of a solid. Despite all of advantages, thermodynamic instability and recrystallization of amorphous pharmaceuticals during the manufacturing or storage is the main reason of restricted use of these materials \cite{25,26}. Thus, understanding molecular mobility of the amorphous drugs to determine their long-term physical stability is essential for the pharmaceutical industry \cite{25,26,22,23,24}. The molecular mobility is characterized by the relaxation processes in the supercooled liquid and glassy states. The structural relaxations, which are irreversible and strongly temperature-dependent, can be experimentally measured using broadband dielectric spectroscopy. A significant growth of the structural relaxation time with cooling can be up to 14 orders of magnitude or more. The processes reveal a coupling between cooperative movements and local dynamics of molecules as a function of temperature in structural rearrangement. In glassy state, the time scale of relaxation is inaccessible to be probed by experiment and simulation. Although the glassy dynamics in amorphous materials has been intensively investigated for many years, there is no currently universal description for the structural relaxation time, the glass transition temperature, and the dynamic fragility. The fundamental mechanisms underlying the phenomenon has remained a challenging problem. 

Glassy dynamics of amorphous materials has been theoretically investigated by the Elastically Collective Nonlinear Langevin Equation (ECNLE) theory. Introduced for the first time by Mirigian and Schweizer \cite{6,10,7,16}, the ECNLE theory is a microscopic and force-based dynamical theory, which views the basic relaxation event as a mixture between the cage-scale (local) hopping process and long range collective motions of the surrounding liquid described by the elastic shoving model \cite{6,10,7,16}. This treatment leads to two distinct, but strongly-related, temperature-dependent barriers. At low temperatures or high densities (deeply supercooled regime),  effects of the elastic barrier on the structural relaxation process become more important and grow much faster when cooling than those of the local cage barrier. The ECNLE theory for rigid molecules associated with a quasi-universal mapping physically understand how the alpha relaxation time depends on temperature over 14 decades \cite{6,10,7,16}. Recall the simulations can only measure relaxation times only over 3-6 decades, and do not access the truly deeply supercooled regime. The thermal mapping allows us to convert a thermodynamic-state-dependent effective hard sphere fluid into a temperature of a real material using experimental equation-of-state data \cite{10}. The theoretical approach has provided both quantitatively and qualitatively good results for  the glass transition temperature, fragility, and the temperature dependence of segmental relaxation time of colloidal suspensions \cite{10,7}, supercooled molecular liquids \cite{6,10} and polymer melts \cite{16}. Recently, Rui and Schweizer have extended the ECNLE theory to investigate the activated relaxation process in binary mixtures \cite{17,18}. However, the extension has not shown a quantitative consistency between theoretical and experimental structural relaxation time. Additionally, without knowledge of the experimental equation-of-state data, it is impossible to determine the temperature dependent structural relaxation time of one- and two-component systems.

In this work, we develop the ECNLE theory by introducing a new thermal mapping model, which is based on a linear relation between the number density of molecules and temperature, to comprehensively interpret the structural relaxation time and glass transition of multi-component amorphous solids. Theoretical calculations for pure materials, and binary and ternary mixtures having different contents are presented. Our numerical results are in quantitative accordance with experimental data.

\section{Theoretical background}
In the framework of ECNLE theory, we model a disordered material as an assembly of a hard sphere single-component system having a volume fraction $\Phi=\rho\pi d^3/6$, where $d$ is the particle diameter and $\rho$ is the number density of particles. The approach considers a single particle motion in a physical picture of slow cage scale dynamics. There are three main forces acting on a tagged particle: (1) the random force, $\delta f$, obeying Gaussian correlations (2) the friction force, $-\zeta_s(\partial r/\partial t)$, here $\zeta_s$ is a short-time friction constant and $r\equiv r(t)$ is the displacement of the particle from its initial position, and (3) the effective caging force, -$\partial F_{dyn}(r)/\partial r$, due to the nearest neighbor interactions confining the particle. $F_{dyn}(r)$ is known as the effective dynamic free energy of the tagged particle caused by the surrounding particles \cite{2,3,4}. Having assumed that the dynamics of the tagged particle in the overdamped limit is governed by the nonlinear stochastic equation \cite{2,3,4}, a force balance equation is
\begin{eqnarray}
-\zeta_s\frac{\partial r}{\partial t} -\frac{\partial F_{dyn}(r)}{\partial r} + \delta f = 0.
\label{eq:1}
\end{eqnarray}
Recall that the thermal noise force obeys the force-force correlation function $\left<\delta f(0)\delta f(t) \right>=2k_BT\zeta_s\delta(t)$, where $k_B$ is the Boltzmann constant and $T$ is temperature. From this, one can derive an analytical expression for the dynamic free energy \cite{6,2,3,4}
\begin{eqnarray}
\frac{F_{dyn}(r)}{k_BT} &=& \int dq\frac{\rho q^2 C^2(q)S^2(q)}{2\pi^2\left[1+S(q)\right]}\exp\left[-\frac{q^2r^2(S(q)+1)}{6S(q)}\right]
\nonumber\\ &-&3\ln\frac{r}{d},
\label{eq:2}
\end{eqnarray}
where $q$ is the wavevector, $S(q)$ is the collective structure factor, and $C(q)=\left[S(q)-1 \right]/\rho S(q)$ is the direct correlation function calculated using Percus-Yevick (PY) integral equation theory \cite{1} for a hard-sphere system. The first term in Eq.(\ref{eq:2}) is responsible for the dynamic mean-field trapping potential and the second term corresponds to the ideal fluid state. 

As the density of the suspension is above a critical value, a barrier free energy, $F_B$, emerges (depicted in Fig. 1) from the effective free energy $F_{dyn}(r)$, which acts as a local barrier to the tagged particle.  The result is that the tagged particle is dynamically arrested by the surrounding particles. We can determine the particle cage radius $r_{cage}$ using the radial distribution function $g(r)$, which relates to the structure factor through $S(q)=1+\rho\int_Vd\mathbf{r}e^{-i\mathbf{qr}}g(r)$. The first minimum position in g(r) is $r_{cage}\approx 1.3-1.5d$. The local minimum and maximum in the free energy profile are the localization length ($r_L \rightarrow \sqrt{\left< (r(t\rightarrow \infty)-r(0))^2 \right>}$) and barrier position ($r_B$). One can obtain the jump distance $\Delta r =r_B-r_L$ from the localization position to the barrier position.

\begin{figure}[htp]
\center
\includegraphics[width=8cm]{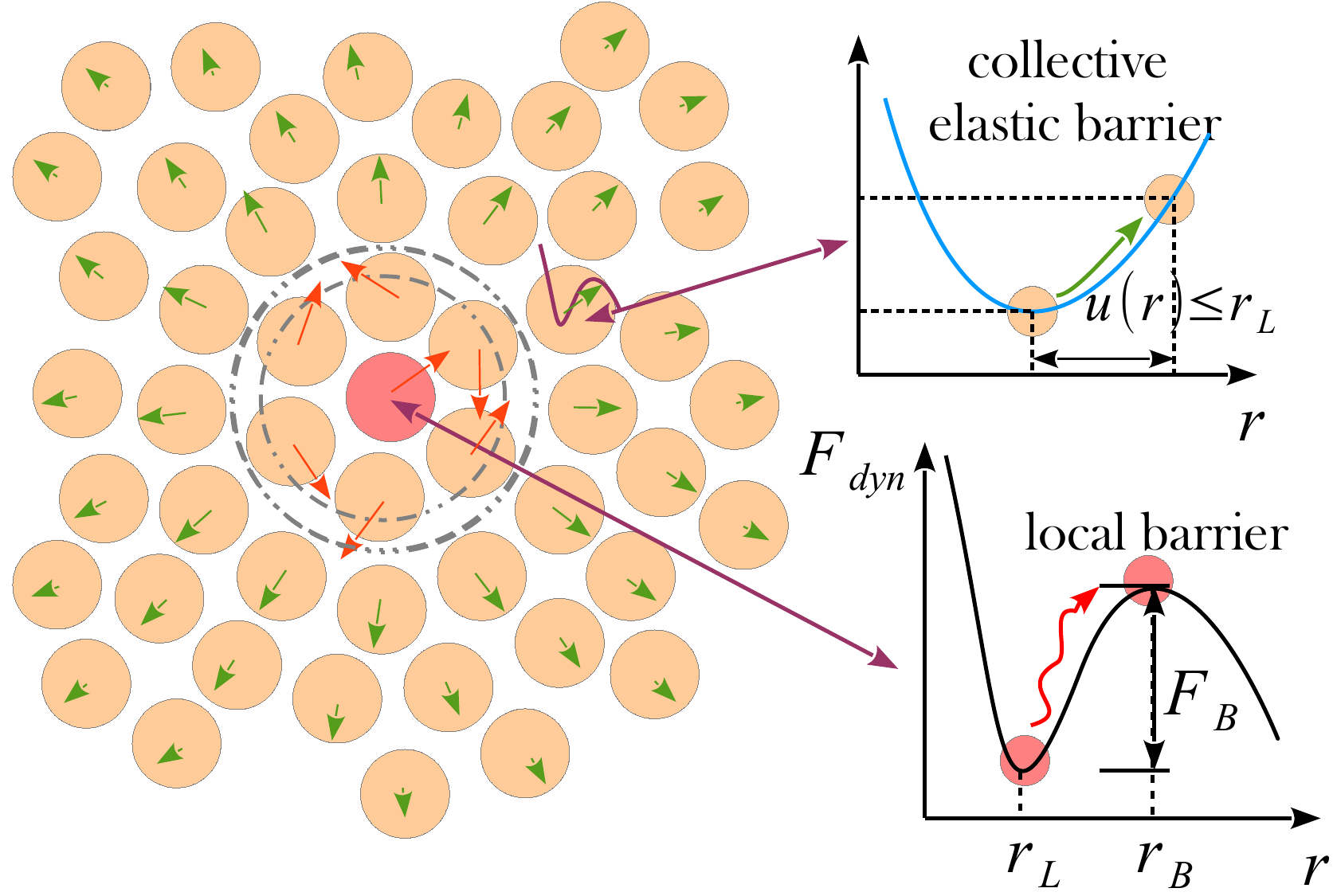}
\caption{\label{fig:1}(Color online) Illustrations of ECNLE theory for structural relaxation process.}
\end{figure}

To capture the collective effects of particles beyond the first shell of the particle cage, we use the shoving model \cite{27} associated with a physical picture of the Einstein model to describe cooperative motions. The re-arrangement of particles in the first shell is required to allow a large hopping, which leads to a small cage dilation on the cage's surface and a harmonic displacement field $u(r)$ of the surrounding medium. The distortion field obeys the elastic continuum mechanics equation \cite{5}
\begin{eqnarray} 
\left(K+\frac{G}{3}\right)\nabla(\nabla.\mathbf{u}) + G\nabla^2\mathbf{u} = 0,
\label{eq:3}
\end{eqnarray}
where $K$ is the bulk modulus and $G$ is the shear modulus. Because of symmetry, the displacement field is a purely radial displacement, and the solution of Eq.(\ref{eq:3}) is found to be
\begin{eqnarray} 
u(r)=Ar+\frac{B}{r^2},
\label{eq:4}
\end{eqnarray}
where $A$ and $B$ are parameters. $A$ is set to be zero to avoid divergence of the strain field at $r\rightarrow \infty$. In Ref.\cite{6,7}, authors found $B=\Delta r_{eff}r_{cage}^2$, where $\Delta r_{eff} \approx 3\Delta r^2/32r_{cage}$ is the cage expansion amplitude. The collective elastic barrier, $F_e$, is estimated by integrating the spatial harmonic displacement energy outside the cage
\begin{eqnarray} 
F_{e} &=& 4\pi\rho\int_{r_{cage}}^{\infty}dr r^2 g(r)K_0\frac{u^2(r)}{2}  \nonumber\\
&\approx& 12\Phi K_0 \Delta r_{eff}^2r_{cage}^3,
\label{eq:5}
\end{eqnarray}
where $K_0 = 3k_BT/r_L^2$ is the spring constant at the localization length as depicted in Fig. \ref{fig:1}. For $r \geq r_{cage}$, $g(r)\approx 1$.

The activated relaxation of particles is characterized by both local and nonlocal processes. Using the Kramer's theory allows us to compute average time for a particle to escape from its particle cage. The time known as the alpha relaxation time, $\tau_\alpha$, is
\begin{eqnarray}
\frac{\tau_\alpha}{\tau_s} = 1+ \frac{2\pi}{\sqrt{K_0K_B}}\frac{k_BT}{d^2} e^{(F_B+F_e)/k_BT},
\label{eq:6}
\end{eqnarray}
where $K_B=\left|\partial^2 F_{dyn}(r)/\partial r^2\right|_{r=r_B}$ is the absolute curvature at the barrier position and $\tau_s$ is a short relaxation time scale. The expression of $\tau_s$ is \cite{6,7}
\begin{eqnarray}
\tau_s=g^2(d)\tau_E\left[1+\frac{1}{36\pi\Phi}\int_0^{\infty}dQ\frac{q^2(S(q)-1)^2}{S(q)+b(q)} \right],
\end{eqnarray}
where $\tau_E$ is the Enskog time scale, $g(d)$ is the coordination number of a particle in system, $b(q)=1/\left[1-j_0(q)+2j_2(q)\right]$, and $j_n(x)$ is the spherical Bessel function of order $n$. Since $\tau_E$ is of the order of $10^{-13}$ s for many thermal liquids and polymers \cite{2,6,7}, for simplicity, we assume $\tau_E = 10^{-13}$ s.    

To quantitatively compare our theoretical calculations with experiments, a thermal mapping from a hard-sphere fluid to a real material is needed. Physically, we propose a thermal mapping model based on the thermal expansion during temperature change. While the material volume linearly expands with temperature, the number of molecules is unchanged. Thus, the number density is 
\begin{eqnarray}
\rho \approx \rho_0\left[1-\beta\left(T-T_{0}\right)\right],
\end{eqnarray}
where $\rho_0$ and $T_{0}$ are the initial number density and temperature, respectively, and $\beta$ is the volume thermal expansion coefficient. Multiplying both sides of the above equation by the volume of a hard sphere particle, $\pi d^3/6$, gives $\Phi \approx \Phi_0\left[1-\beta\left(T-T_0\right)\right]$. Then, we can deduce
\begin{eqnarray}
T \approx T_0 - \frac{\Phi - \Phi_0}{\beta\Phi_0}.
\label{eq:7}
\end{eqnarray}
The above linear relationship between temperature and a volume fraction can be obtained by fitting the original thermal mapping of Mirigian and Schweizer \cite{2,6,7} for a variety of polymers and thermal liquids having a wide range of dynamic fragility. In a recent work \cite{2}, Phan and Schweizer showed $\Phi_0 = 0.5$ and $\beta\Phi_0 = 6\times 10^{-4}$ $K^{-1}$. Note that the linear thermal expansion coefficient of many supercooled liquids have similar values $2-5\times 10^{-4}$ $K^{-1}$ \cite{31,32}. Thus, the volume thermal expansion coefficient $\beta$ is in the range of $6-15 \times 10^{-4}$ $K^{-1}$ and one has $\beta\Phi_0=3-7.5 \times 10^{-4}$ $K^{-1}$. Consequently, in the present work, we use this value $\beta\Phi_0 = 6\times 10^{-4}$ $K^{-1}$ for all amorphous drugs since it is very reasonable to experiments and consistent with the original thermal mapping \cite{2,6,7}. 

Now all physics of molar mass and particle size are embedded in the characteristic parameter $T_0$. Since the structural relaxation time $\tau_\alpha$ is approximately 100 $s$ at $\Phi \approx 0.61$ in our calculations, we can estimate the characteristic temperature $T_0$ for a given pure material via its experimental value of $T_g$, which is defined by $\tau_\alpha(T=T_g) = 100s$. Note that the original thermal mapping introduced by Mirigian and Schweizer \cite{10} is based on the temperature-dependent reproduction of a density fluctuation amplitude (compressibility) in the long wavelength limit for thermal liquids, $S_0=\rho k_BT\kappa_T$. The dimensionless compressibility carried out by the PY theory \cite{1} for hard sphere fluids, $S_0^{HS}$, is equal to the experimental data. The mapping relation is
\begin{eqnarray}
S_0^{HS}&=&\frac{(1-\Phi)^4}{(1+2\Phi)^2}\equiv S_{0,exp}=\rho k_BT\kappa_T\nonumber\\
&\approx &\frac{1}{N_s}\left(-A_1+\frac{B_1}{T} \right)^{-2},
\label{eq:8}
\end{eqnarray}
where $A_1$ and $B_1$ are parameters corresponding to the molecular level packing contribution and cohesive energy in equation of state (EOS), respectively, and $N_s$ is the interaction site number in a rigid molecule. Although quantitative values of $T_g$ given by the original thermal mapping for several polymers are not identical to experimental values, plotting $\tau_\alpha$ versus $T_g/T$ provides a good agreement with experiments \cite{10,16}, and the shift of average glass transition temperature over polymer films describes experiments and simulations \cite{11} well. However, the thermal mapping cannot be obtained without the experimental EOS data.

\section{Experimental section}
\subsection{Materials and ternary system preparation}
Crystalline forms of ezetimibe (EZB) and simvastatin (SVS) of purity greater than 99 $\%$ and molecular mass equal to $M_{wEZB}$ = 409.4 g/mol, and $M_{wSIM}$ = 418.6 g/mol were purchased from Polpharma (Starogard Gdański, Poland). EZB and SIM are described chemically as ((3R,4S)-1-(4-fluorophenyl)-3-[(3S)-3-(4-fluorophenyl)-3-hydroxypropyl]-4-(4-hydroxyphenyl) azetid in-2-one) and butanoic acid, 2,2-dimethyl-, (1S,3R,7S,8S,8aR)-1,2,3,7,8,8a-hexahydro-3,7-dimethyl-8-[2-[(2R,4R)-tetrahydro-4-hydroxy-6-oxo-2H-pyran-2-yl]ethyl]-1-naphthalenyl ester, respectively. Kollidon VA64 polymer (KVA) of molecular mass $M_w = 45000-47000$ g/mol was purchased from BASF SE (Germany) and used as recieved. Ternary amorphous system of EZB, SVS, and KVA having a weight ration of 0.2:0.2:0.6 was prepared by the quench cooling technique. To acquire homogeneous system we mixed (prior the vitrification) all components at appropriate ratios in mortar for approximately 15-20 min.

\subsection{Differential Scanning Calorimetry}
Thermodynamic properties of amorphous form of EZB, SVS, KVA, binary system of EZB 1:1 SVS, and ternary system containing EZB 1:1 SVS + 60wt. $\%$ KVA were examined by means of a Mettler-Toledo differential scanning calorimetry (DSC) 1 STARe System. The instrument was equipped with an HSS8 ceramic sensor having 120 thermocouples. The measuring device was calibrated for temperature and enthalpy using indium and zinc standards. Samples were measured in an aluminum crucible (40 $\mu L$). All measurements were carried out with a heating rate equal to 10 $K$/min.

\subsection{Broadband Dielectric Spectroscopy}
We investigate molecular mobility of EZB, SVS, KVA, binary EZB 1:1 SVS system, and ternary system of EZB-SVS-KVA using a Novo-Control GMBH Alpha dielectric spectrometer. Dielectric spectra were measured in the broad frequency range $10^{-1}$ Hz - $10^6$ Hz during heating from $T \sim T_g$ up to 416, 356, 473, 373, and 407 $K$ for EZB, SVS, KVA, the binary and ternary system, respectively. During dielectric experiments we controlled the temperature by a Quatro temperature controller with temperature stability better than 0.1 $K$. Prior the broadband dielectric spectroscopy (BDS) experiments the samples were quench cooled on the heating plate. The examined systems were measured in a parallel-plate cell made of stainless steel (diameter of 15 mm, and a 0.1 mm gap provided by silica spacer fibers). The dielectric loss data have been fitted using the Havrilak - Negami (HN) function to determine the temperature dependence of $\tau_\alpha$ of the examined materials. The empirical HN approach is defined as follows
\begin{eqnarray}
\varepsilon^*_{HN}(\omega)=\varepsilon' (\omega) - i\varepsilon'' (\omega)=\varepsilon_\infty + \frac{\Delta \varepsilon}{\left[1+(i\omega\tau_{HN})^a \right]^b},
\end{eqnarray}
where $\varepsilon'$ and $\varepsilon''$ are real and imaginary parts of complex dielectric function, $\varepsilon_\infty$ in the above equation denotes high frequency limit permittivity, $\Delta \varepsilon$ represents dielectric strength, $\omega = 2\pi f$, $\tau_{HN}$ is the HN relaxation time, while $a$ and $b$ are the symmetric and asymmetric broadening of relaxation peak. The fitting parameters obtained from the presented above formula were then used to calculate the structural relaxation time using the equation
\begin{eqnarray}
\tau_\alpha = \tau_{HN}\left[\sin\left(\frac{\pi a}{2+2b} \right) \right]^{-1/a}\left[\sin\left(\frac{\pi ab}{2+2b} \right) \right]^{1/a}.
\end{eqnarray}

\section{Numerical results and discussions}
\begin{figure}[htp]
\center
\includegraphics[width=8.5cm]{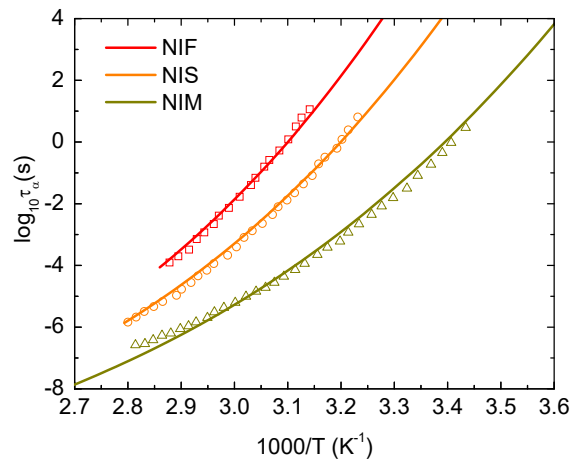}
\caption{\label{fig:2}(Color online) Logarithm of the bulk relaxation time as a function of inverse normalized temperature for nifedipine, nisoldipine, and nimodipine. The solid curves and data points correspond to theoretical calculations and experimental data, respectively.}
\end{figure}

Figure \ref{fig:2} shows the theoretical and experimental temperature dependence of the structural alpha relaxation times for nifedipine (NIF), nisoldipine (NIS), and nimodipine (NIM). A quantitatively good agreement between our theory and previous experiment \cite{8} is obtained, with relatively perfect overlaping. The glass transition temperature, $T_g$, for NIF, NIS, and NIM, carried out by BDS techniques, are 315, 305, and 285 K, respectively \cite{8}. From these values, one can use Eq. (\ref{eq:7}) to find $T_{0,NIF} \approx 498$ $K$, $T_{0,NIS} \approx 488$ $K$, and $T_{0,NIM} \approx 470$ $K$ for our thermal mapping. The derivation of $T_0$ guarantees the experimental and theoretical $T_g$ values are identical. The consistency between the theoretical and experimental thermal variation is expected as a heritance of the success of the original mapping \cite{10}. 

Now to apply our theoretical model to binary mixtures, motivated by an empirical Gordon-Taylor equation \cite{12} for $T_g$ of two-component composites, we propose 
\begin{eqnarray}
T_0=\frac{w_1T_{0,1}+w_2T_{0,2}}{w_1+w_2},
\label{eq:9}
\end{eqnarray}
where $T_0$ and $T_{0,i}$ are characteristic temperatures for binary
mixtures and their pure components, and $w_1$ and $w_2$ are weight proportions of each component in the composites. Since the Gordon-Taylor equation provides good predictions for $T_g$ of binary mixtures in Ref.\cite{8,13} and $T_0$ is proportional to $T_g$, one can expect to find an appropriate value $T_0$ for composite system. Here we suppose that the interaction between the components is hard-sphere or there is no complex chemical interaction between the components.

\begin{figure}[htp]
\center
\includegraphics[width=8.5cm]{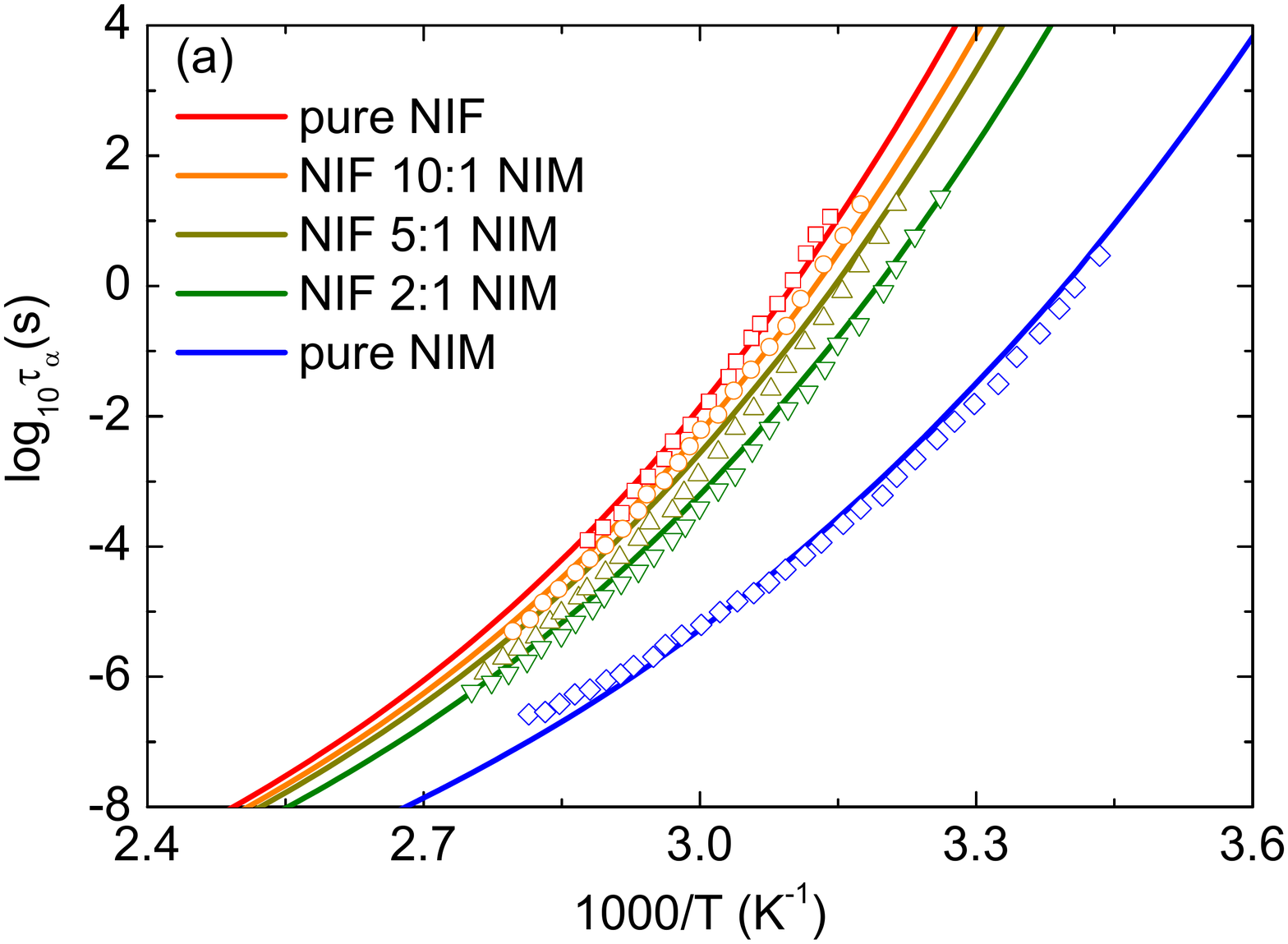}
\includegraphics[width=8.5cm]{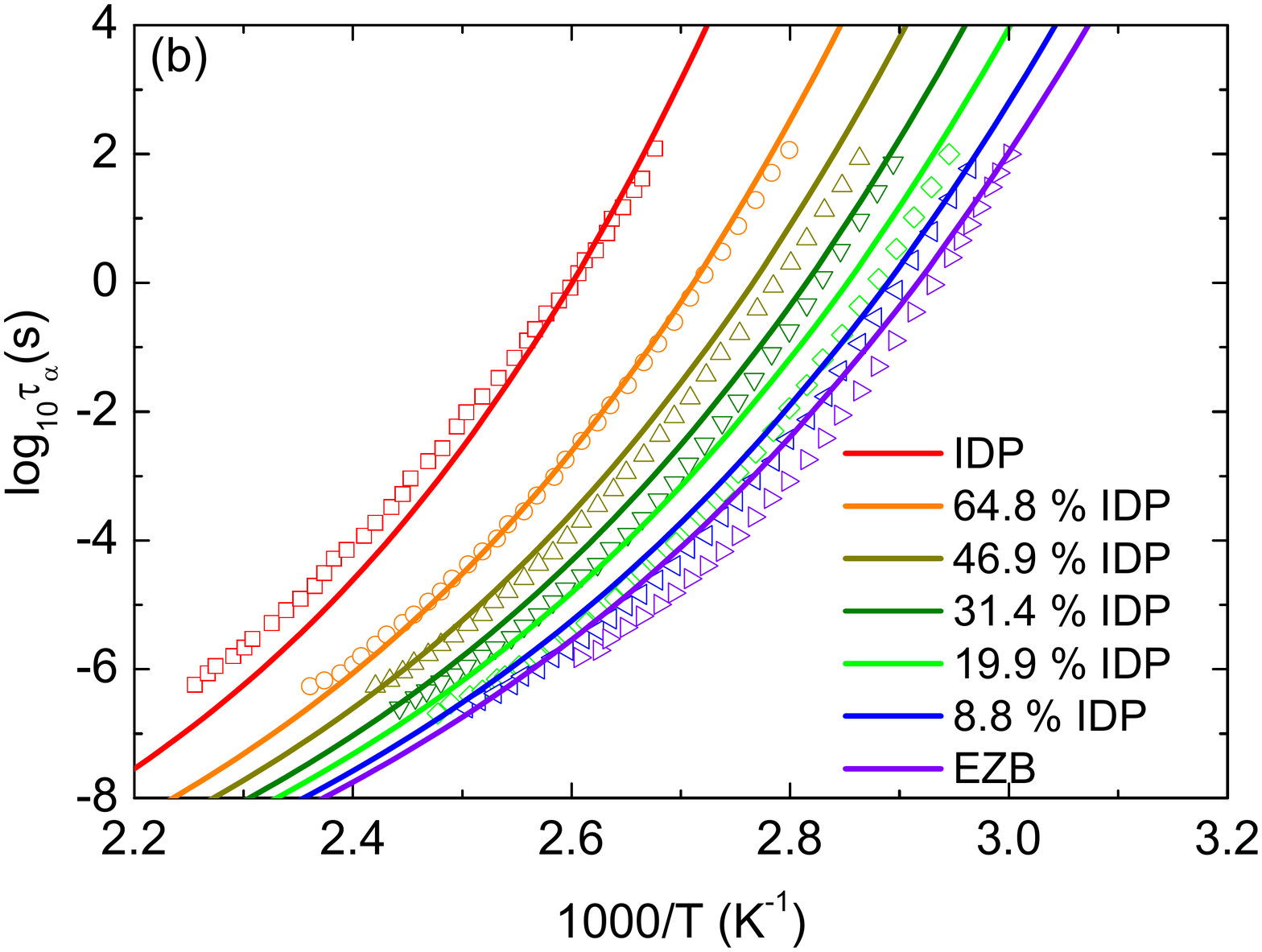}
\caption{\label{fig:3}(Color online) Logarithm of the alpha relaxation time versus inverse normalized temperature for (a) pure NIF and NIM, and binary NIF-NIM mixtures having various weight proportions of 10:1, 5:1, and 2:1, and (b) pure EZB and IDP, and binary EZB-IDP mixtures having various weight ratios of IDP: 0.088, 0.199, 0.314, 0.469, and 0.648. The solid curves and data points correspond to theoretical calculations and experimental data, respectively.}
\end{figure}

Figure \ref{fig:3}a presents the alpha relaxation times for binary systems of NIF/NIM for several weight fractions as a function of normalized inverse temperature over a wide range of time scales. Our theoretical calculations are in quantitatively good accordance with the prior experimental results \cite{8}. This finding suggests that the proposed relation (Eq. (\ref{eq:9})) for the parameter $T_0$ of the binary mixtures of NIF/NIM works well. However, recall that the molecular structures of NIF and NIM are analogous. To validate the generality of Eq. (\ref{eq:9}), testing the relation for two-component composites having different chemical structures is necessary. This is a reason why we have to exhibit other data sets for non-analogous drugs, as depicted in Fig. \ref{fig:3}b.

Figure \ref{fig:3}b shows the theoretical and experimental structural relaxation times as a function of temperature for pure amorphous EZB,  indapamide (IDP), and their binary mixtures with several weight fractions of IDP. Our theoretical calculations agree quantitatively with the experimental data in Ref. \cite{13}. Since the experimental glass transition temperatures of EZB and IDP are 333 and 373.5 $K$, respectively, one obtains $T_0 = 518$ and 558.5 $K$ for EZB and IDP, respectively. The discrepancies between theory and experiment in Fig. \ref{fig:3}b are greater than those in Fig. \ref{fig:3}a, as seen by the naked eye. There are two possible reasons for the deviation. First, the chemical structures of EZB and IDP drug are different, and $\tau_s$ cannot be identical. Recall that we set $\tau_E = 0.1$ $ps$ for all amorphous materials in our calculations. Second, some chemical interactions between these two drugs may occur. However, to zero-order approximation, the relatively good theory-experiment consistency suggests that Eq.(\ref{eq:9}) gives good predictions for the temperature-dependent $\tau_\alpha$ of various co-amorphous materials.

\begin{figure}[htp]
\center
\includegraphics[width=8.5cm]{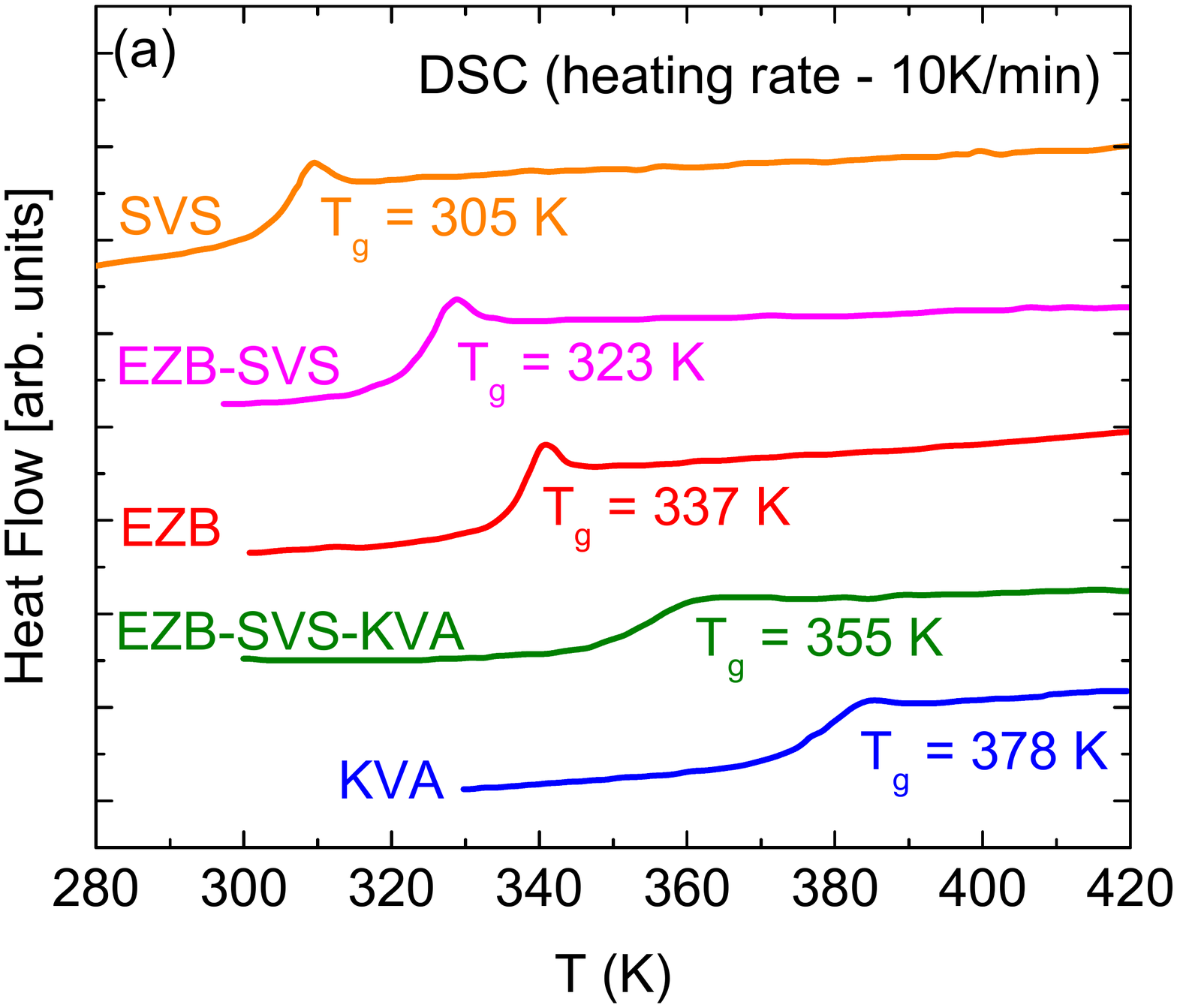}
\includegraphics[width=8.5cm]{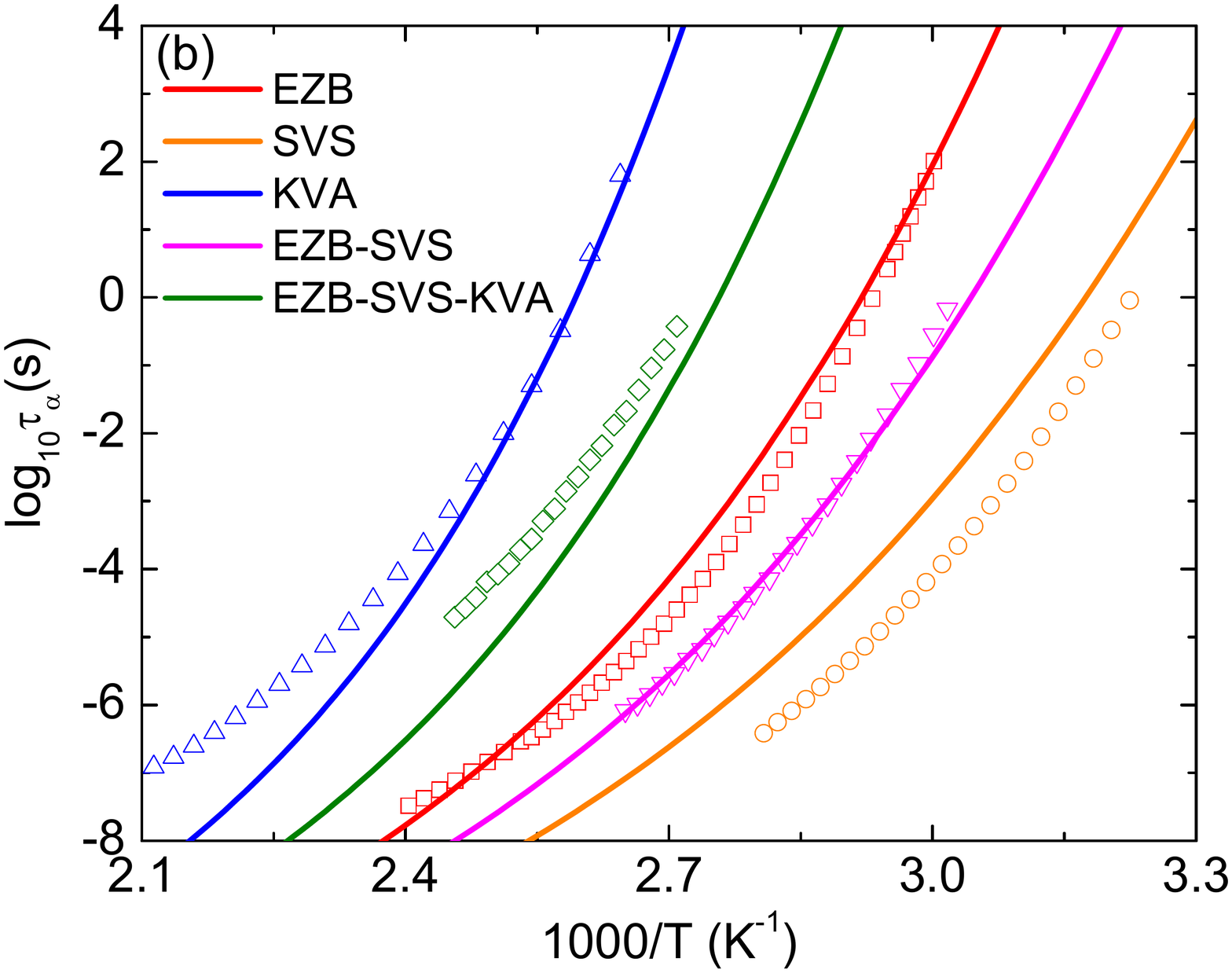}
\caption{\label{fig:4}(Color online) (a) DSC thermograms with heating rate 10 $K$/min and (b) logarithm of the structural relaxation time versus inverse normalized temperature for  pure EZB, SVS, and KVA, binary EZB-SVS system having a weight ratio of 1:1, and ternary EZB-SVS-KVA mixtures with the weight ratio of 0.2:0.2:0.6. The solid curves and data points correspond to theoretical calculations and BDS experimental data, respectively.}
\end{figure}

Now to verify feasible applications of Eq. (\ref{eq:9}) to ternary mixtures, we carry out experiments for the structural relaxation times as a function of $1000/T$ for pure EZB drug, pure SVS drug, pure KVA polymer, two-component EZB-SVS mixtures with the weight ratio of 1:1, and their composite EZB-SVS-KVA having the weight ratio of 0.2:0.2:0.6. The obtained DSC traces with heating rate of 10 $K$/min are presented in Fig. \ref{fig:4}a. For each system a single glass transition event was registered indicating samples homogeneity. Glass transition temperatures were determined as the midpoint of the heat capacity increment. $T_g$ values obtained from DSC experiments are collected in Fig. \ref{fig:4}a. It was determined that in the supercooled liquid region the temperature evolution of the structural relaxation time of the investigated materials shows a non-Arrhenius behavior and can be well parameterized by the Vogel-Fulcher-Tammann (VFT) equation:
\begin{eqnarray}
\log_{10}\tau_\alpha(T)=\log_{10}\tau_\infty+\frac{DT_{VFT}}{T-T_{VFT}},
\end{eqnarray}
where $\log_{10}\tau_\infty$, $T_{VFT}$, and $D$ are fitting parameters. 

Using the VFT fitting parameters we calculate value of the glass transition temperature defined as a temperature at which $\tau_\alpha = 100$ s. The obtained values are in a perfect agreement with the $T_g$s calculated using the DSC  method (see Table \ref{tab:table1}).

\begin {table*}[htp]
\caption{\label{tab:table1} The VFT fitting parameters and the $T_g$ values obtained from the DSC measurements, and the $T_g$ value obtained from the BDS measurements.}
\begin{center}
\begin{tabular}{|c|c|c|c|c|c|}
\hline
 & EZB/SVS & SVS & EZB & EZB/SVS/KVA & KVA  \\
\hline
$\log_{10}\tau_\infty$ & -13.79 $\pm$ 0.19 & -15.68 $\pm$ 0.13 & -16.30 $\pm$ 0.15 & -13.65 $\pm$ 0.46 & -11.62 $\pm$ 0.08  \\
\hline
$DT_{VFT}$ & 1823 $\pm$ 67 & 2386 $\pm$ 51 & 2663 $\pm$ 9 & 2404 $\pm$ 203 & 1584 $\pm$ 31 \\
\hline
$T_{VFT}$ (K) & 273.1 $\pm$ 1.3 & 244 $\pm$ 0.9 & 270 $\pm$ 0.2 & 290.2 $\pm$ 3.9 & 326.8 $\pm$ 0.8 \\
\hline
$T_g$ BDS (K) & 323 & 303 & 333 & 356 & 378  \\
\hline
$T_g$ DSC (K) & 323 & 303 & 336 & 355 & 378 \\
\hline
\end{tabular}
\end{center}
\end {table*}

Comparisons between our theoretical calculations with experimental data are shown in Fig. \ref{fig:4}b. Using the glass transition temperatures in Table \ref{tab:table1} and Eq. (\ref{eq:7}) gives us $T_0 = 490.5$ $K$ and 561 $K$ for SVS and KVA, respectively. While the numerical results of EZB drug nearly overlap experimental data over 12 decades in time, the theoretical structural relaxation time for KVA polymer exhibits perfect consistency with the experiment at low temperatures, where $\log_{10}\tau_\alpha$ ranges approximately from -3 to 2.  Although our calculations for SVS drug describe qualitatively well the thermal variation of the experimental counterpart, the deviation is quantitatively noticeable. These calculations are still encouraging since up to now there is no adjustable parameter needed. Interestingly, theoretical predictions for the temperature dependence of $\log_{10}\tau_\alpha$ against experiment for  the binary EZB-SVS mixture over nearly 6 orders of magnitude of measured relaxation time reveal good accordance.


We then extend Eq.(\ref{eq:9}) to determine the characteristic temperature for the three-component composite
\begin{eqnarray}
T_0=0.2T_{0,EZB}+0.2T_{0,SVS}+0.6T_{0,KVA}.
\label{eq:10}
\end{eqnarray}

The theoretical curve for EZB-SVS-KVA composite is quite close to its experimental data. The main reasons why our approach works relatively well are open and challenging problems. However, small quantitative disagreement is predictable and unsurprising. One can raise a question of whether a phase separation in the amorphous ternary system might lead to the deviation of our theoretical calculations from experimental results. Recall that our DSC measurements shown in Fig. \ref{fig:4}a clearly evidence that the three-component system is homogeneous. Consequently, the experiment-theory discrepancy is not related to emergence of phase separation. It is likely caused by the simplicity of our thermal mapping when ignoring biological and chemical complexities in reality, and structural changes due to conformational connectivity and penetration. The mapping captures physics of translational motions of molecules and totally neglects all rotational motions. Additionally, elementary molecules are supposed to be impenetrable. Any violation of the assumptions has substantial influences on the temperature dependence of $\tau_\alpha$. 

Despite some approximations made in the present approach, our extended ECNLE theory provides opportunities to investigate activated events below $T_g$. In a prior work \cite{16}, Mirigian and Schweizer convincingly presented that experimental segmental relaxation times for polypropylene, polybutadiene, and polyvinylacetate are well-described by the original ECNLE theory, particularly below $T_g$ where $\tau_\alpha$ spans from 100 s to $10^6$ s. This agreement suggests physical interpretation of two-barrier activated relaxation is still valid to the smooth growth of theoretical relaxation time with cooling below $T_g$, but apply to amorphous materials having no distinctive separation (dynamic decoupling) from structural relaxation near glass transition temperature. Recent experiments have revealed that the dynamic decoupling in the relaxation process at low temperatures ( $< Tg$) can be found in some drugs such as bicalutamide \cite{29} and celecoxib \cite{28}. This interesting problem is currently under study. From our perspectives, the dynamic decoupling is likely due to the temperature dependence of thermal expansion, which is strongly related to $\alpha_t$ in our thermal mapping, in amorphous materials. 

\section{Conclusions}
We have proposed a new approach to investigate the temperature dependence of alpha structural relaxation times for pure amorphous drugs and their composites with different weight ratios. The approach is based on the ECNLE theory applied to the hard sphere fluid associated with the new thermal mapping from hard sphere density to real materials. A key parameter for the thermal mapping of a one-component system is calculated from its the glass transition temperature. Then, the parameters of different single materials associated with weight ratios in their composite are used to estimate this parameter for the multiple-component composite. Our theoretical calculations quantitatively agree with a variety of experiments. The approach would enable us to step forward in understanding the structural relaxation times of co-amorphous drugs and their physical stability.
\begin{acknowledgments}
This research is funded by Vietnam National Foundation for Science and Technology Development (NAFOSTED) under Grant No. 103.01-2016.61. The author, M. P., is grateful for the financial support received within the Project No. 2015/16/W/NZ7/00404 (SYMFONIA 3) from the National Science Centre, Poland. This work was supported by JSPS KAKENHI Grant Numbers JP19F18322 and JP18H01154.
\end{acknowledgments}

\end{document}